\documentclass[aps,prb,twocolumn,letterpaper,superscriptaddress,showpacs]{revtex4-2}
\usepackage{graphicx}
\usepackage{CJK}
\usepackage{verbatim}
\usepackage{physics}
\usepackage[colorlinks,linkcolor=blue,anchorcolor=blue, citecolor=blue,urlcolor=blue,]{hyperref}
\usepackage{lineno}
\usepackage{bm}
\usepackage{mathptmx}
\makeatletter
\newcommand{\parallelsum}{\mathbin{\!/\mkern-5mu/\!}}
\begin{document}
\begin{CJK*}{UTF8}{bsmi}
\title{Strongly correlated doped hole carriers in the superconducting nickelates:
Their location, local many-body state, and low-energy effective Hamiltonian}
\author{Zi-Jian Lang (\CJKfamily{gbsn}郎子健)}
\affiliation{Tsung-Dao Lee Institute \& School of Physics and Astronomy, Shanghai Jiao Tong University, Shanghai 200240, China}
\author{Ruoshi Jiang (\CJKfamily{gbsn}姜若诗)}
\affiliation{Tsung-Dao Lee Institute \& School of Physics and Astronomy, Shanghai Jiao Tong University, Shanghai 200240, China}
\author{Wei Ku (\CJKfamily{bsmi}顧威)}
\altaffiliation{corresponding email: weiku@sjtu.edu.cn}
\affiliation{Tsung-Dao Lee Institute \& School of Physics and Astronomy, Shanghai Jiao Tong University, Shanghai 200240, China}
\affiliation{Key Laboratory of Artificial Structures and Quantum Control (Ministry of Education), Shanghai 200240, China} 

\date{\today}

\begin{abstract}
The families of high-temperature superconductors recently welcomed a new member: hole doped nickelate Nd$_{0.8}$Sr$_{0.2}$NiO$_2$ with a $\sim$15K transition temperature.
To understand its emergent low-energy behaviors and experimental properties, an immediate key question is whether the superconducting hole carriers reside in oxygen as in the cuprates, or in nickel as in most nickelates.
We answer this crucial question via a ``(LDA+$U$)+ED'' scheme: deriving an effective interacting Hamiltonian of the hole carriers from density functional LDA+$U$ calculation, and studying its local many-body states via exact diagonalization.
Surprisingly, distinct from the expected Ni$^{2+}$ spin-triplet state found in most nickelates, the local ground state of two holes is actually a Ni-O spin-singlet state with second hole greatly residing in oxygen.
The emerged eV-scale model therefore resembles that of the cuprates, advocating further systematic experimental comparisons.
Tracing the microscopic origin of this unexpected result to the lack of apical oxygen in this material, we proposed a route to increase superconducting temperature, and a possible quantum phase transition absent in the cuprates.

\end{abstract}

\maketitle
\end{CJK*}

Despite intensive research efforts since its discovery three decades ago in cuprates~\cite{Muller}, high-temperature superconductivity remains one of the most important puzzles in the field of condensed matter physics~\cite{Dagotto}.
For example, the behaviors of the superconducting gap~\cite{Shi2008,Damascelli}, dominant phase fluctuation~\cite{Bozovic,Emery1995}, and properties under magnetic field~\cite{Ando1999,Giraldo2018,Zavaritsky2002} all demonstrate novel characteristics qualitatively distinct from standard superconductors\cite{Bardeen}.
In fact, it is now well documented~\cite{Dagotto,Lee2006} that even above the superconducting temperature, the ``normal'' state of the cuprates is also qualitatively different from the textbook Fermi liquid state.
To give a few examples, the occurrence of a pseudogap~\cite{Damascelli}, linear resistivity~\cite{Dagotto,Legros2019}, non-saturating high-temperature resistivity~\cite{Takenaka2003}, sign-changing Hall coefficient~\cite{Tsukada2006}, optical conductivity demonstrating continuum and mid-infrared peak~\cite{Hwang2007}, nematic and inhomogeneous charge distribution~\cite{Yazdani2013,Wise2008}, together establish strongly that at sub-eV scale the cuprates are in an unknown quantum state of matter, whose proper description is still lacking.

The challenge to formulate a proper understanding of this atypical state of the matter lies in the strong correlation between the hole carriers in this system, whose complexity grows at lower energy and becomes intractable below 100meV.
Nevertheless, rather reasonable understanding has been achieved at the 1eV scale via extensive studies of the charge~\cite{Phillips2005}, antiferromagnetic~\cite{Anderson1950} and polaronic~\cite{Alexandrov} correlations.
Specifically, it is well-established~\cite{Chen1991,Zannen} that the hole carriers reside mainly in the oxygen atoms and strongly entangle with spins of the intrinsic holes in copper atoms~\cite{Anderson1987,Zhang,Emery} within 1eV scale, and double occupation of the copper atoms can be ``integrated out'' in the low-energy theory.
These physical effects heavily dress the carriers in any low-energy theory below this 1eV scale as in, for example, the $t$-$J$ model~\cite{Spatek1977}.

To gain further control at the sub-eV scale for such a complex problem of emergent physics, having multiple similar families for comparison is obviously highly valuable.
This makes the recent discovery of superconducting Nd$_{0.8}$Sr$_{0.2}$NiO$_2$ nickelate with a $\sim$15K transition temperature~\cite{Li} extremely exciting.
Unlike the tetrahedral-coordinated, multi-orbital Fe-based superconductors discovered a decade ago~\cite{Kamihara}, Nd$_{0.8}$Sr$_{0.2}$NiO$_2$ has similar point-group symmetry to the cuprates and the same valence electron count (nine electrons per transition metal atom).
One can thus hope for valuable insights from thorough comparison with the cuprates in all aspects of measured properties.

One immediate key issue is the location of the hole carriers in superconducting Nd$_{0.8}$Sr$_{0.2}$NiO$_2$ nickelate, or more precisely its dominant local many-body state at low energy.
In typical nickel oxides, one expects that the large Hund's coupling would favor adding holes to the Ni$^{+}$ ion to form Ni$^{2+}$ spin-triplet states $\ket{d_{x^2-y^2\uparrow}d_{3z^2-r^2\uparrow}}$ of the holes~\cite{Sawatzky2019,Jiang,Zhang2020,Chang,Hu,Ashvin2019}.
In that case, one should be able to ``integrate out'' the oxygen degrees of freedom at sufficiently low energy.
However, if the hole carriers turn out to greatly reside in the oxygen atoms instead, this would not be possible and \emph{a qualitatively different low-energy physics} (kinetics, interactions, and correlations) \emph{would emerge} involving a different set of dominant many-body states~\cite{Wu,Mei}.
In fact, this is exactly what happens in the cuprates, in which one integrates out the double occupation of holes in Cu instead, and the doped holes form a local Cu-O spin-singlet state $\ket{\text{Cu}_\uparrow \text{O}_\downarrow}-\ket{\text{Cu}_\downarrow \text{O}_\uparrow}$ with the intrinsic hole in Cu$^{2+}$ ion~\cite{Anderson1987,Zhang}.
Correspondingly, the low-energy carriers are under a very strong on-site charge and neighboring spin-correlation that dictate all the unconventional behaviors of the cuprates~\cite{note}.

From this perspective, it is essential to address the key question concerning the location of the hole carriers in this material.
However, current experimental conclusions are quite confusing.
Recent x-ray absorption data~\cite{Hepting2020} was interpreted as support for the Kondo-lattice model or Anderson-lattice model, in which the itinerant Nd electrons are able to screen the Ni local moments~\cite{Zhang2020}.
If this interpretation is correct, even in hole-doped samples, the doped holes would go to the Nd atoms and remove the electron carriers.
This should reduce their screening and lead to an increasing average magnetic moment.
This is, however, \emph{opposite} to the recent magnetization measurement~\cite{Wen2020}, which shows a reducing average moment upon doping.

Therefore, it is of ultimate importance and timeliness at this stage of the field to settle this key issue of location of hole carriers in this family, both theoretically and experimentally.
Only then, would one be able to move on to the following important questions concerning this family:
(1) What are the dominant local many-body states associated with the hole carriers?
(2) What is the eV-scale effective Hamiltonian of the hole carriers?
(3) What are the similarities and differences to the cuprates?
(4) Is there possible new physics absent in the cuprates?

Here, we address all of these essential questions via a (LDA+$U$)+ED scheme: examining the energies of the local many-body eigenstates of a realistic high-energy \emph{interacting} Hamiltonian, extracted from density functional calculation of NdNiO$_2$ within the LDA+$U$ approximation~\cite{Liechtenstein,Anisimov}.
Surprisingly, due to the strong kinetic coupling between Ni and O, the hole carriers are found to reside largely in the \emph{oxygen} and form a Ni-O spin-singlet state with the intrinsic holes in Ni$^{+}$ ion, similar to the case in the cuprates.
We then derive the effective eV-scale theory that reveals its similarity and difference to the cuprates, advocating systematic studies contrasting with the cuprates in revealing essential factors of the outstanding puzzle of the high-temperature superconductivity.
Furthermore, we trace our unexpected result that differs from the usual suspect Ni$^{2+}$ spin-triplet state in the typical nickelates, to the absence of apical oxygen in this family.
This therefore opens up the possibility of a quantum phase transition absent in the cuprates via introduction of apical oxygen or uni-axial pressure.

\begin{figure}
	\begin{center}
		\includegraphics[width=7.5cm]{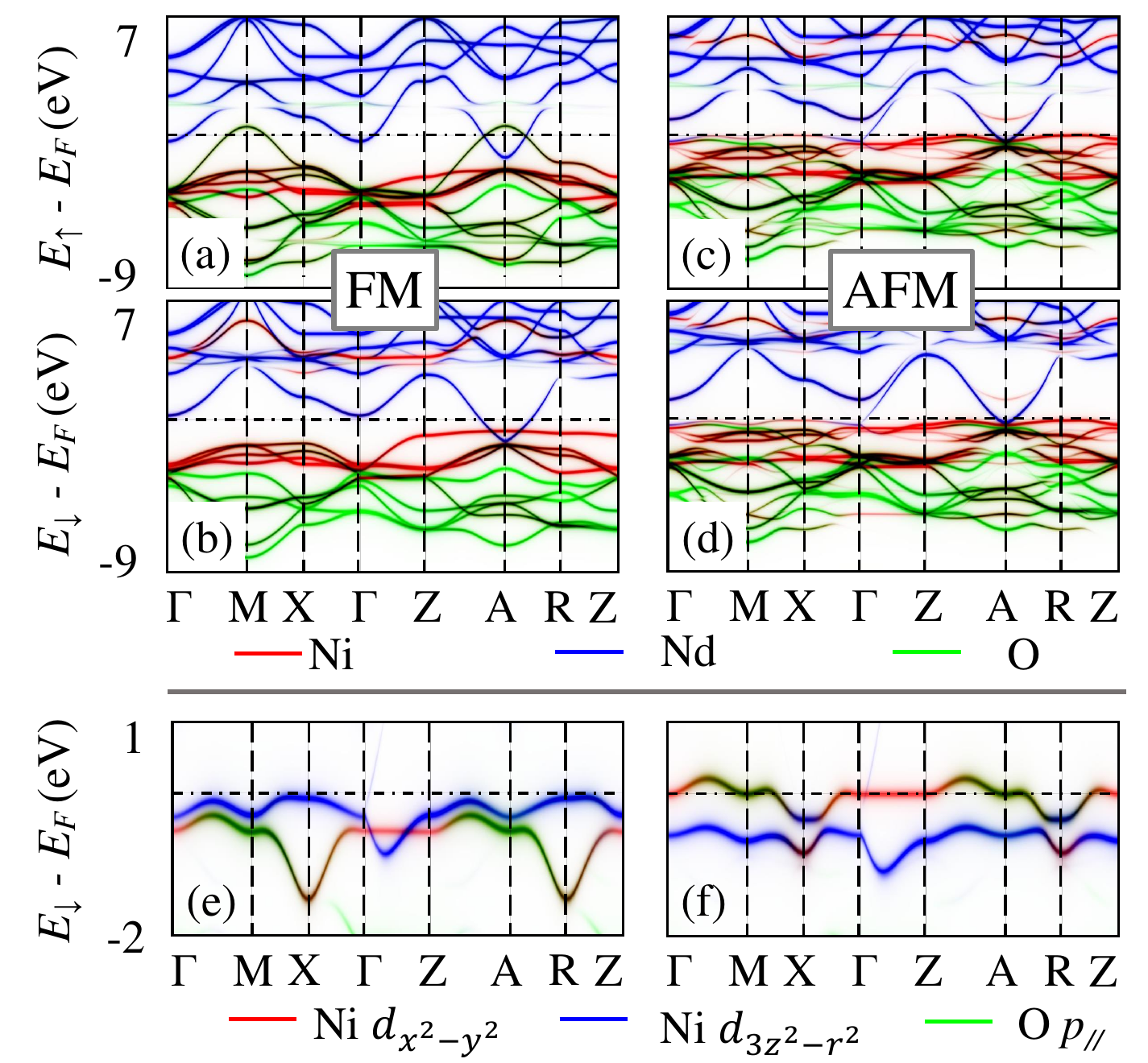}
	\vspace{-0.3cm}
	\caption{Spin majority [(a)(c)] and minority [(b)(d)] band structure of undoped NdNiO$_2$ under FM [(a)(b)] and AFM [(c)(d)] orders from LDA+$U$ calculations, presented (unfolded) in the one-Ni Brillouin zone with relative weight of Ni, Nd, and O atoms shown in red, blue, and green.
	Note that spin-majority aligned Nd-$f$ orbitals are made transparent to provide a clearer picture, which breaks weakly the symmetry between panels (c) and (d). Detailed comparison near the Fermi energy for AFM ordered undoped NdNiO$_2$ (e) and 25\% doped Nd$_3$SrNi$_4$O$_8$(f) shows that the doped holes reside mainly in O-$p_{\parallelsum}$ (green) and Ni-$d_{x^2-y^2}$ (red) orbitals.
	} 
	\vspace{-0.9cm}
	\label{fig1}
	\end{center}
\end{figure}

Let's first consider the most relevant high-energy orbital space of Nd$_{0.8}$Sr$_{0.2}$NiO$_2$ based on our density functional calculation within the LDA+$U$ approximation~\cite{Liechtenstein,Anisimov,supplementary} with $U-J=0.6-0.06$ Ry.
Figures.~\ref{fig1}(a) and (b) show our resulting band structure of the undoped NdNiO$_2$ parent compound with ferromagnetic (FM) order.
The single unoccupied red Ni band in panel (b) makes it clear that Ni is indeed in a Ni$^+$ ($d^9$) configuration containing an intrinsic hole in the spin-minority channel.
Detailed analysis further indicates that the intrinsic holes reside in the Ni-$d_{x^2-y^2}$ orbital.
In addition, in the FM cases, one observes a clear electron pocket from a blue Nd band across the Fermi energy around A$=(\pi,\pi,\pi)$, accompanied by additional holes in a Ni-O hybrid band.
This self-doping effect has been reported previously~\cite{supplementary,Lee,Botana2020,Choi2020} and is considered an important aspect of the electronic structure of this material~\cite{Sawatzky2019,Zhang2020}.
However, an unfrustrated square lattice of spin-1/2 nickel should host an intrinsic strong anti-ferromagnetic (AFM) correlation, with or without a long-range order, and thus this issue might be better examined via an AFM order.
For an easier comparison, Figs.~\ref{fig1}(c) and (d) show the AFM band structure unfolded~\cite{Wei2010} in the one-Ni unit cell.
With strong AFM correlation, the electron pocket now shrinks to a negligibly small size, and nearly no self-doped electrons reside in the Nd atoms. 

Now, where would the hole carriers reside?
Figure~\ref{fig1}(e) highlights the orbitals that contribute to the LDA+$U$ band structure near the Fermi energy.
It suggests that the low-energy hole carriers are likely to occupy the (blue) spin-minority Ni-$d_{3z^2-r^2}$ orbital and form a Ni$^{2+}$ ion, consistent with the intuition from the typical nickelates~\cite{Sawatzky}.
Surprisingly, the same analysis in a doped Nd${_3}$SrNi$_4$O$_8$ system [c.f.Fig.~\ref{fig1}(f)] shows instead that the hole carriers actually reside in a hybrid of O-$p_{\parallelsum}$ (green) and spin-majority Ni-$x^2-y^2$ (red) orbitals, a scenario more commonly seen in doped charge-transfer insulators like the cuprates.
Such a dramatic contrast is obviously related to the strong local Coulomb interaction that requires a more accurate treatment than LDA+$U$, to be discussed below in detail.

Interestly, our Wannier orbital analysis~\cite{supplementary, Wei2002, Wei2006} shows that the orbital spaces hosting the low-energy electron and hole carriers couple very weekly ($<0.7$eV) to each other compared to their large energy separation ($>6.8$eV)
Consequently, all virtual-process enabled many-body (kinetic, super-exchange, or electron-hole annihilating) couplings are at best 10meV scale or smaller.
In addition, with the Nd-$f$ local moments set only to be along the global spin-majority direction, the Ni and O orbitals in our AFM results in Figs.~\ref{fig1}(c) and (d) still show negligible spin-dependence, indicating negligible spin-spin coupling to the Ni and O orbitals.
Therefore up-to at most 10\% inaccuracy, the Hamiltonians describing these two types of carriers decouple from each other and can be considered separately.
We will therefore drop the Nd orbitals from our investigation of the local physics of the hole carriers in this study.

We thus proceed to construct~\cite{supplementary, Wei2002, Wei2006, Liechtenstein, Slater} a realistic interacting Hamiltonian, from which low-energy local many-body states of holes can be carefully examined.
With a filling factor of 1+$x$ (one intrinsic hole plus $x$ low-energy hole carriers), the following high-energy ($\sim$10eV scale) four-orbital interacting Hamiltonian of \emph{holes} contains only the four most relevant low-energy degrees of freedom, $x^2-y^2$ and $3z^2-r^2$ orbitals of Ni and two $p_{\parallelsum}$ orbitals pointing toward Ni from the two O sites next to Ni:
\begin{equation}
    \begin{split}
        H^{\text{eff}} &= \sum_{i,m,\nu}\epsilon_{im}n_{im\nu}+\sum_{i,i^\prime,m,m^\prime,\nu}t_{ii^\prime mm ^\prime}c^\dagger_{im\nu}c_{i^\prime m^\prime \nu}\\
        &+\sum_{<j,i>m} J_m\mathbf{S}_j^F\cdot\mathbf{S}_{im} \\
        &+\sum_{i,m1,m2,m3,m4,\nu,\nu^\prime} U_{m1,m2,m3,m4,\nu,\nu^\prime} c^\dagger_{im1\nu}c^\dagger
        _{im2\nu^\prime}c_{im3\nu^\prime}c_{im4\nu},
    \end{split}
    \label{eq1}
\end{equation}
where $c^\dagger_{im\nu}$ and $n_{im\nu}\equiv c^\dagger_{im\nu}c_{im\nu}$ denote the standard creation and number operator of \emph{holes} of spin $\nu$ in orbital $m$ of site $i$ that experience kinetic hopping $t_{ii^\prime mm ^\prime}$ and orbital energy $\epsilon_{im}$.
Only effective on-site repulsive interactions $U_{dd^\prime}$ and Hund's coupling $J_{dd^\prime}$ among holes in the Ni-$d$ orbitals are considered, since the O-$p$ orbitals are mostly unoccupied or singly occupied by holes.
For completeness, orbital dependent coupling to the large 3/2 spin $S^F$ of nearest Nd-$f$ orbitals at location $j$ is included as well, even though its effect is of no significance in our analysis below.
(Here $\mathbf{S}_{im}=\sum_{\nu,\nu^\prime}c^\dagger_{im\nu}\bm{\sigma}_{\nu,\nu^\prime}c_{i m \nu^\prime}$ denotes the orbital-dependent spin operator and $\bm{\sigma}_{\nu,\nu^\prime}$ is the vector of Pauli matrices.
The use of LDA+$U$ here allows a more accurate orbital space in constructing effective interacting Hamiltonians.
All parameters are available in the Supplementary Material~\cite{supplementary}.)

\begin{figure}
	\begin{center}
	\includegraphics[width=7.8cm]{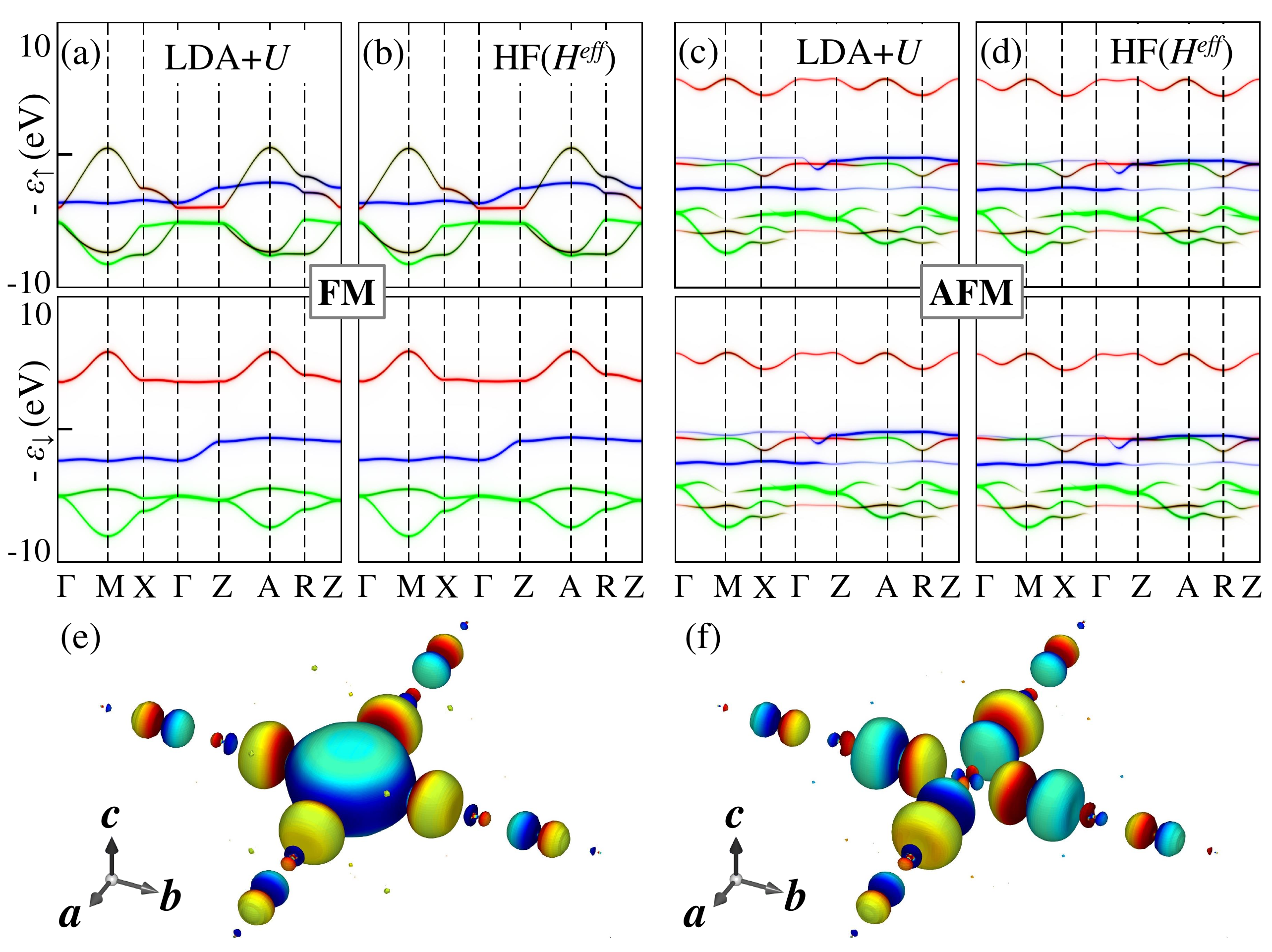}
	\end{center}
	\vspace{-0.8cm}
	\caption{Comparison of the Hartree-Fock band structures of our four-orbital model [(b), (d)] and those of LDA+$U$ ones [(a), (c)] within the same Ni $d_{x^2-y^2},d_{3z^2-r^2}$ and O $p_{\parallelsum}$ subspace, under FM [(a), (b)] and AFM [(c), (d)] order.
	(e), (f) Illustration of outer-shell Wannier orbitals for symmetric $P_s$(e) and anti-symmetric $P_a$(f)  superposition of O $p_{\parallelsum}$ orbitals, colored according to the density gradient, from negative (blue) to positive (red).}
	\label{fig2}
	\vspace{-0.6cm}
\end{figure}

Figure~\ref{fig2} demonstrates an excellent correspondence between the LDA+$U$ band structure within the subspace of the selected orbitals, and our model band structure under the same Hartree-Fock treatment of local interactions.
Notice, especially, that the agreement occurs under both FM and AFM orders with the \emph{same} set of parameters, despite the significant difference in these two orders.
These results thus establish the high quality of these parameters and the validity of our model Hamiltonian in describing various magnetic structures, including the paramagnetic parent compound.

To further investigate the local many-body state, it is convenient to represent the four-orbital space by a set of \emph{Ni-centered orthonormal} orbitals, such that the system is reduced into a simple tetragonal lattice of only Ni sites~\cite{Zhang}.
This allows a simple definition of local many-body space associated with these orbitals at the same Ni site: two inner-shell orbitals Ni-$d_{x^2-y^2}$, Ni-$d_{3z^2-r^2}$, and two outer-shell orbitals $P_a$ and $P_s$ as anti-symmetric and symmetric superposition of O-$p$ orbitals~\cite{Wei2013} shown in Figs.~\ref{fig2}(e) and (f).
(Technically, this is achieved by exploiting the gauge freedom~\cite{Vanderbilt} of the O-$p$ Wannier orbitals~\cite{Wei2013}.)

Table~\ref{tab1} gives our resulting local kinetic parameters for the holes.
As expected, the inner-shell $d_{x^2-y^2}$ orbital has the lowest energy, and thus naturally hosts one intrinsic hole.
Naturally, local symmetry of the orbital dictates that the $d_{x^2-y^2}$ orbital couples strongly ($\sim$2.56eV) to the $P_a$ in the outer shell, and $d_{3z^2-r^2}$ much more weakly ($\sim$0.69eV) to $P_s$.
Notice that the outer shell (O-$p$ derived orbitals) is about 8eV above the inner shell (Ni-$d$ orbitals), which is of comparable size to $U_{dd}$.
This explains the strong tendency toward charge transfer of the hole carriers: It costs similar energy to add a hole to the inner-shell orbitals or to the outer-shell orbitals.

\begin{table} 
    \caption{Local on-site and kinetic energy in $H^{eff}$ for inner-shell $d_{x^2-y^2}$, $d_{3z^2-r^2}$, and outer-shell $P_a$, $P_s$ orbitals.}
    \begin{ruledtabular}
        \begin{tabular}{lllll}
$H^{eff}$(eV)              & $d_{x^2-y^2}$ & $P_a$ & $d_{3z^2-r^2}$ & $P_s$ \\ \hline
$d_{x^2-y^2}$  & -4.09        & -2.56              & 0  &0       \\
$P_a$ & -2.56             & 4.10         &0       & 0 \\
$d_{3z^2-r^2}$          & 0          & 0              & -3.51     & 0.69     \\
$P_s$          & 0         & 0              & 0.69     & 6.19    
\end{tabular}
    \end{ruledtabular}
    \label{tab1}
	\vspace{-0.4cm}
\end{table}

Now we can obtain a more accurate many-body picture via exact diagonalization (ED) of Eq.~\ref{eq1} within the above local many-body space containing two holes.
We find the local ground state to be a Ni-O spin-singlet state $\ket{\text{Ni}_\uparrow \text{O}_\downarrow}-\ket{\text{Ni}_\downarrow \text{O}_\uparrow}$ involving mostly one hole in $d_{x^2-y^2}$ and another hole in $P_a$-$d_{x^2-y^2}$ hybrid, similar to the well-known Zhang-Rice singlet in the cuprates~\cite{Zhang}.
It has slight lower energy than the next competitive state, a Ni$^{2+}$ spin-triplet state $\ket{d_{x^2-y^2\uparrow}d_{3z^2-r^2\uparrow}}$, by $\sim$0.1eV.
Upon inclusion of the remaining kinetic energy across sites ($\sim$1.2eV, to be discussed below), this energy difference can easily approach an eV (consistent with the recent RIXS measurement~\cite{Hepting2020}), given the much weaker kinetic energy ($\sim$0.4eV) of the Ni$^{2+}$ spin-triplet state.
This large energy scale renders our conclusion insensitive to the choice of local interactions~\cite{supplementary}.

This somewhat unexpected result actually explains qualitatively the recent observation of reduced magnetic moments against doping~\cite{Wen2020}.
Since the local magnetic moment of a spin-singlet (spin 0) is smaller than the parent compound (spin-1/2), hole doping should weaken the averaged magnetic moment and the associated magnetization.
The Ni$^{2+}$ spin-triplet (spin-1) state popularly believed to be the local ground state, however, would enhance the magnetization and contradict with recent observations~\cite{RIXSnote}.
We expect more experimental observations on the magnetic and electronic properties to further confirm our findings.

From a phenomenological consideration of high-temperature superconductivity, our finding also makes strong physical sense.
To sustain thermal fluctuation at a temperature comparable to the cuprates, it is necessary to have strong enough kinetic energy to establish correspondingly large phase stiffness.
The large in-plane spread of the Ni-O spin-singlet state produces the necessary large kinetic energy (bandwidth $\sim2.4$eV~\cite{supplementary}), comparable to that of the cuprates.
In contrast, Ni$^{2+}$ spin-triplet state hosts a much smaller kinetic energy ($\sim0.7$eV in our estimation~\cite{supplementary}) along the out-of-plane direction only.
It is really hard to imagine that such weak quasi-1D kinetics are able to resist phase fluctuations at a temperature scale similar to the cuprates.

\begin{table} 
    \caption{Comparison of estimated model parameters of $H^{\it{1}B}$ for Nd$_{1-x}$Sr$_{x}$NiO$_2$ and La$_{2-x}$Sr$_x$CuO$_4$~\cite{Ogata2008} in eV}.
    \begin{ruledtabular}
        \begin{tabular}{lllll}
  & $t$  & $t^\prime$ & $t^{\prime\prime}$ & $J$  \\ \hline
Nd$_{1-x}$Sr$_{x}$NiO$_2$ & 0.31 & -0.07       & 0.06                  & 0.07 \\
La$_{2-x}$Sr$_x$CuO$_4$   & 0.40      & -0.07          &  0                  & 0.13
\end{tabular}
    \end{ruledtabular}
    \label{tab2}
	\vspace{-0.3cm}
\end{table}

With these analyses, we can finally proceed to make a crude estimation of a low-energy ($\sim$eV scale) effective one-band Hamiltonian of holes within the many-body subspace in which each site is either singly occupied, $\ket{1_i}$, or doubly occupied, $\ket{s_i}$, by holes in the Ni-O spin-singlet state~\cite{Spatek1977,Yin2010,tnote}.
Given the similarity to the cuprates in the transition-metal lattice and in the singlet formation of hole carriers, it is not surprising to find our resulting Hamiltonian for holes,
 \begin{equation}
    \begin{split}
        H^{{\it1}B}=\sum_{ii^\prime\nu}t_{ii^\prime}\tilde{c}_{i\nu}^\dagger \tilde{c}_{i^\prime\nu} + \sum_{<i,j>}J\mathbf{S}_i\cdot\mathbf{S}_j,
     \end{split}
    \label{H_eff}
\end{equation}
resembles the well-studied $t$-$J$ model of the cuprates if one ignores the weak coupling to the Nd-$f$ local spins.
Here, $\tilde{c}_{i\nu}^\dagger$ creates a dressed hole at site $i$ of spin $\nu$ under the constraint that each site must be at least singly occupied.
Similarly, the spin operator $\mathbf{S}_i=\sum_{\nu,\nu^\prime}\tilde{c}^\dagger_{i\nu}\bm{\sigma}_{\nu,\nu^\prime}\tilde{c}_{i\nu^\prime}$ is subject to the same constraint.
Table~\ref{tab2} shows our estimation of the hopping parameters $t_{ii^\prime}$ from $\mel{s_i 1_{i^\prime}}{H^{\text{eff}}}{1_i s_{i^\prime}}$, and exchange parameter $J$ via $\frac{4t^4}{\Delta^2U_{dd}}+\frac{8t^4}{\Delta^2(2\Delta+U_{pp})}$ using charge transfer gap $\Delta\approx6$eV~\cite{deltaNote}, on-site repulsion $U_{dd}=8.8$eV~\cite{supplementary} for $d_{x^2-y^2}$, and $U_{pp}=4$eV~\cite{Ogata2008} for O-$p$ orbitals.

Our analysis above provides some clear indication of how this high-temperature superconductor family compares with the well-studied but still unsolved cuprates.
At the above eV scale, the different chemical nature of Ni and Cu atoms renders this family quite different from the cuprates.
An essential aspect is the quantitatively weaker involvement of oxygen orbitals in hosting the hole carriers owing to Ni-$d$ orbitals' higher chemical energy.
This would surely be reflected in the spectral distribution of most experimental spectroscopes at the 10eV scale.

Interestingly, at sub-eV scale, owing to the strong Coulomb interaction, Table~\ref{tab2} shows that other than a slightly smaller energy scale than the cuprates, the emergent low-energy Hamiltonian of this nickelate family resembles the cuprates extremely well.
It is therefore not surprising that besides the unconventional superconductivity, this family also demonstrates many unusual and puzzling low-energy behaviors~\cite{Li} observed in the cuprates~\cite{Takenaka2003}, for example, the lack of saturation of high-temperature resistivity (so-called ``bad metal'' behavior) and the linear temperature dependence of low-temperature resistivity (so-called ``strange metal'' behavior).
As more experimental observations are conducted, we expect more reports of resemblance in the low-energy electronic structure and physical behavior.

Our further analysis indicates that the lack of apical oxygen above the Ni atoms in this structure has a positive influence on the stability of the spin-singlet state.
In the absence of the O (outer-shell) orbitals, the ground state of our local exact diagonalization becomes the Ni$^{2+}$ spin-triplet state, as expected from Hund's rule.
Therefore, the stability of the Ni-O spin-singlet state benefits greatly from the virtual kinetic coupling between the Ni-$d_{x^2-y^2}$ orbitals and the O-$P_a$ orbitals.
In other words, the lack of apical oxygen in this structure really helps push the energy of the Ni$^{2+}$ spin-triplet state higher by (1) raising the crystal-field energy of $d_{3z^2-r^2}$ by $\sim$0.6eV (in hole picture, c.f. Table~\ref{tab1}), and (2) losing the virtual kinetic energy involving oxygen orbitals.
It also removes the ``super-repulsion'' effects~\cite{Wei2010} that might suppress superconductivity.
One can thus expect a higher superconducting transition temperature upon application of a in-plane pressure.

On the other hand, owing to the relatively higher chemical energy of $d$ orbitals in Ni, our study suggests rich possibilities of new physics absent in the cuprates.
Specifically, we anticipate a local many-body level crossing from the above Ni-O spin-singlet state to the Ni$^{2+}$ spin-triplet state, via introduction of apical oxygen or $c$-axis pressure tuning of the structure.
This opens up an exciting possibility of quantum phase transition from a cuprate-like system into a high-spin (likely magnetic) system.
Such an additional tunability implies a much richer physics in this nickelate family, careful investigations of which might help clarify the essential underlying physics of the normal and superconducting state of these unconventional superconductors.

In conclusion, employing a (LDA+$U$)+ED scheme, we investigate the timely issue regarding the location (and corresponding local many-body state) of hole carriers in the recently discovered Nd$_{0.8}$Sr$_{0.2}$NiO$_2$ nickelate unconventional superconductors that determines the appropriate modeling of the eV- (and lower) scale physics.
We find that the hole carriers reside mostly in the O and spin-majority Ni-$d_{x^2-y^2}$ orbitals, as opposed to the expected spin-minority Ni-$d_{3z^2-r^2}$ orbital commonly found in the nickelates.
Employing a local many-body calculation of two holes, we find that the second hole forms a local Ni-O spin-singlet with the intrinsic hole, similar to that in the cuprates, while the previously expected Ni$^{2+}$ spin-triplet state is located almost an eV higher in energy.
This puts this nickelate family in a similar category to the cuprates, but nicely expands the parameter space.
Our conclusion is in good agreement with the magnetization measurement and makes perfect sense to the cuprate-like transport properties that demonstrates similarly strange metal and bad metal behaviors.
Our analysis also suggests that the sensitivity to influences of apical oxygen allows for the tuning of superconducting temperature.
It should also open up additional exciting possibilities of quantum phase transition absent in the cuprates.
Altogether, with such a close contrast to the cuprates, studies of this nickelate family are expected to reveal essential clues for the long-standing puzzle of unconventional superconductivity and non-Fermi liquid behavior of the cuprates, two of the most important issues in modern condensed matter physics.
\begin{acknowledgments}
We thank Hai-Hu Wen and Jie Ma for useful discussions.  This work is supported by National Natural Science Foundation of China (NSFC) \#11674220 and 11745006 and Ministry of Science and Technology \#2016YFA0300500 and 2016YFA0300501.
\end{acknowledgments}
\bibliography{MainTex.bib}

\begin{thebibliography}{58}
\bibitem [{tno()}]{tnote}%
  \BibitemOpen
  \href@noop {} {}\bibinfo {note} {Virtual processes to the Ni$^{2+}$ triplet
  state can introduce additional corrections to our resulting 2nd and 3rd
  neighboring hoppings beyond this crude estimation. Yet, these longer-range
  hoppings are much weaker than the nearest neighboring one by almost an order
  of magnitude, so they should not alter the qualitative aspects of our
  discussion below.}\BibitemShut {Stop}%
\end{thebibliography}
\end{document}